\documentclass[superscriptaddress,amsmath,amssymb,aps,pre,twocolumn,footinbib]{revtex4}

\usepackage[normal]{subfigure}
\usepackage{amsmath}
\usepackage{dcolumn}
\usepackage{bm}

\usepackage{pxfonts}

\usepackage[normalem]{ulem}
\usepackage{graphicx,color}
\usepackage{subfigure}
\usepackage{bbold}
\usepackage{cancel}
\usepackage[colorlinks,citecolor=red,linkcolor=blue]{hyperref}
\usepackage[usenames,dvipsnames,svgnames]{xcolor}
\usepackage{calligra}
\DeclareMathAlphabet{\mathcalligra}{T1}{calligra}{m}{n}

\newcommand{\ee}{\end{equation}}
\newcommand{\be}{\begin{equation}}

\newmuskip\pFqmuskip

\newcommand*\pFq[6][8]{%
  \begingroup 
  \pFqmuskip=#1mu\relax
  \mathcode`\,=\string"8000
  \begingroup\lccode`\~=`\,
  \lowercase{\endgroup\let~}\pFqcomma
  {}_{#2}F_{#3}{\left[\genfrac..{0pt}{}{#4}{#5};#6\right]}%
  \endgroup
}
\newcommand{\pFqcomma}{\mskip\pFqmuskip}

\begin{document}
\title{Interfacial tension and a three-phase generalized self-consistent theory of non-dilute soft composite solids}

\author{Francesco Mancarella}
\affiliation{Nordic Institute for Theoretical Physics
(NORDITA), SE-106 91 Stockholm, Sweden}
\author{Robert W. Style}
\affiliation{Mathematical Institute, University of Oxford, Oxford OX1 3LB, UK}
\author{John S. Wettlaufer}
\affiliation{Yale University, New Haven, Connecticut 06520, USA}
\affiliation{Mathematical Institute, University of Oxford, Oxford OX1 3LB, UK}
\affiliation{Nordic Institute for Theoretical Physics
(NORDITA), SE-106 91 Stockholm, Sweden}
\begin{abstract}
In the dilute limit Eshelby's inclusion theory captures the behavior of a wide range of systems and properties.  However, because Eshelby's approach neglects interfacial stress, it breaks down in soft materials as the inclusion size approaches the elastocapillarity length $L$. Here, we use a three-phase generalized self-consistent method to calculate the elastic moduli of composites comprised of an isotropic, linear-elastic compliant solid hosting a spatially random monodisperse distribution of spherical liquid droplets. As opposed to similar approaches, we explicitly capture the liquid-solid interfacial stress when it is treated as an isotropic, strain-independent surface tension.  Within this framework, the composite stiffness depends solely on the ratio of the elastocapillarity length $L$ to the inclusion radius $R$.  Independent of inclusion volume fraction, we find that the composite is stiffened by the inclusions whenever $R < 3L/2$. Over the same range of parameters, we compare our results with alternative approaches (dilute and Mori-Tanaka theories that include surface tension). Our framework can be easily extended to calculate the composite properties of more general soft materials where surface tension plays a role. 
\end{abstract}
\date{\today}
\maketitle

\section{Introduction}
Composite materials are of interest because they rarely have the bulk properties of their constituents alone.  Thus, understanding their properties provides a challenge and a test bed for both controlling material behavior and understanding how and why natural materials have evolved \cite[e.g.,][and refs. therein]{Alain:2015,Bertoldi:2015}. Because in both engineering and natural settings there is always a compromise between the ability to tailor every detail and achieving an optimal effective behavior, such as stiffness, theoretical approaches that span the widest range of key control parameters are desirable. 

Among the most successful idealized geometric models for two-phase matrix-inclusion composites is Hashin's composite-spheres model \cite{Hashin62, Hashin64}, where the actual composite is replaced by  a set of ``composite spheres'' with a suitable size-distribution, and arranged in a volume-filling configuration. Each composite sphere consists of a homogeneous sphere representing the inclusion phase, surrounded by a concentric spherical shell of matrix material.  The ratio between internal and external radii of each shell is determined in terms of the volume fraction occupied by the inclusion phase within the {\em actual} composite.  While the effective bulk modulus is determined analytically,  there is no exact solution for the effective shear modulus, although considerable information is available on its variational bounds.

In addition to the bounds mentioned above, a broad class of self-consistent (SC) methods have been developed that yield  analytical predictions for both bulk and shear moduli.  For example, Kroner \cite{Kroner58} introduced a self-consistent approximation wherein the inclusion itself is directly embedded in an unknown homogeneous effective medium.  Budiansky \cite{Budiansky65} and Hill \cite{Hill65a,Hill65b} use this approach in their model for elastic moduli of composites, although they noted physical inconsistencies at high inclusion volume fractions $\phi$.  

A three-phase generalized self-consistent (GSC) model was introduced by Kerner \cite{Kerner56, footnote1} and van der Poel \cite{Pol58}.  Christensen and Lo \cite{Christensen79} took this approach by replacing the set of all actual inclusions by a single ideal inclusion, in the
 ``composite spheres''  framework described above.  Whereas the SC method is generally simpler than GSC models, the proper boundary conditions of the latter remove the unphysical behavior of the former as $\phi$ becomes large.  Both approaches typically assume continuity of strain (or no slip) across interfaces \cite[e.g.,][]{Shick94}, although the bulk modulus has been shown to be unaffected by finite slip  \cite{Takahashi78}. 

The surface stress at solid/liquid interfaces can have a substantial range of size-dependent effects in soft materials. For example, recent work has shown that surface stress significantly influences pearling and creasing instabilities \cite{Mora10, Mora11, Chakrabarti13,Henann14}, wetting  \cite{Style12,StylePRL13,StylePNAS13,Nadermann13,Bostwick14,Karpitschka14}, adhesion \cite{StyleNatComm13,Salez13,Xu14,Cao14}, and the relaxation of soft solids towards their equilibrium shapes \cite[e.g.,][]{Mora13}.  
Hence, here we aim to reformulate the micromechanics of soft composites in the non-dilute limit to include the effects of surface stress.  In so doing, we systematically examine how the interplay between the inclusion volume fraction $\phi$ and the inclusion size $R$ influences the mechanical properties.   We note that, for both dilute and non-dilute cases, the inclusion/matrix surface stress has been treated in previous work, the most notable of which however assumes linear (in-strain) surface stress \cite{Duan05,Duan07,Duan07secondo,Brisardbulk10, Brisard10}, and/or use incorrect boundary conditions, as described previously \cite{Stylesoft15}.

Style et al., \cite{Style15, Stylesoft15} studied a dilute monodisperse random spatial distribution of liquid droplets of radius $R$ embedded in a homogeneous isotropic elastic solid matrix.  They included the effect of surface tension at the inclusion/matrix interface.  Here, we use a three-phase GSC model to examine the stiffness in such a composite system for finite inclusion volume fractions.  We compare the results with the dilute theory and an extension of the Mori-Tanaka theory to non-dilute soft composite solids \cite{MSWa}, both of which include surface tension.  

\section{The model}
\begin{figure}[htp]
\centering 
\includegraphics[width=0.9\columnwidth]{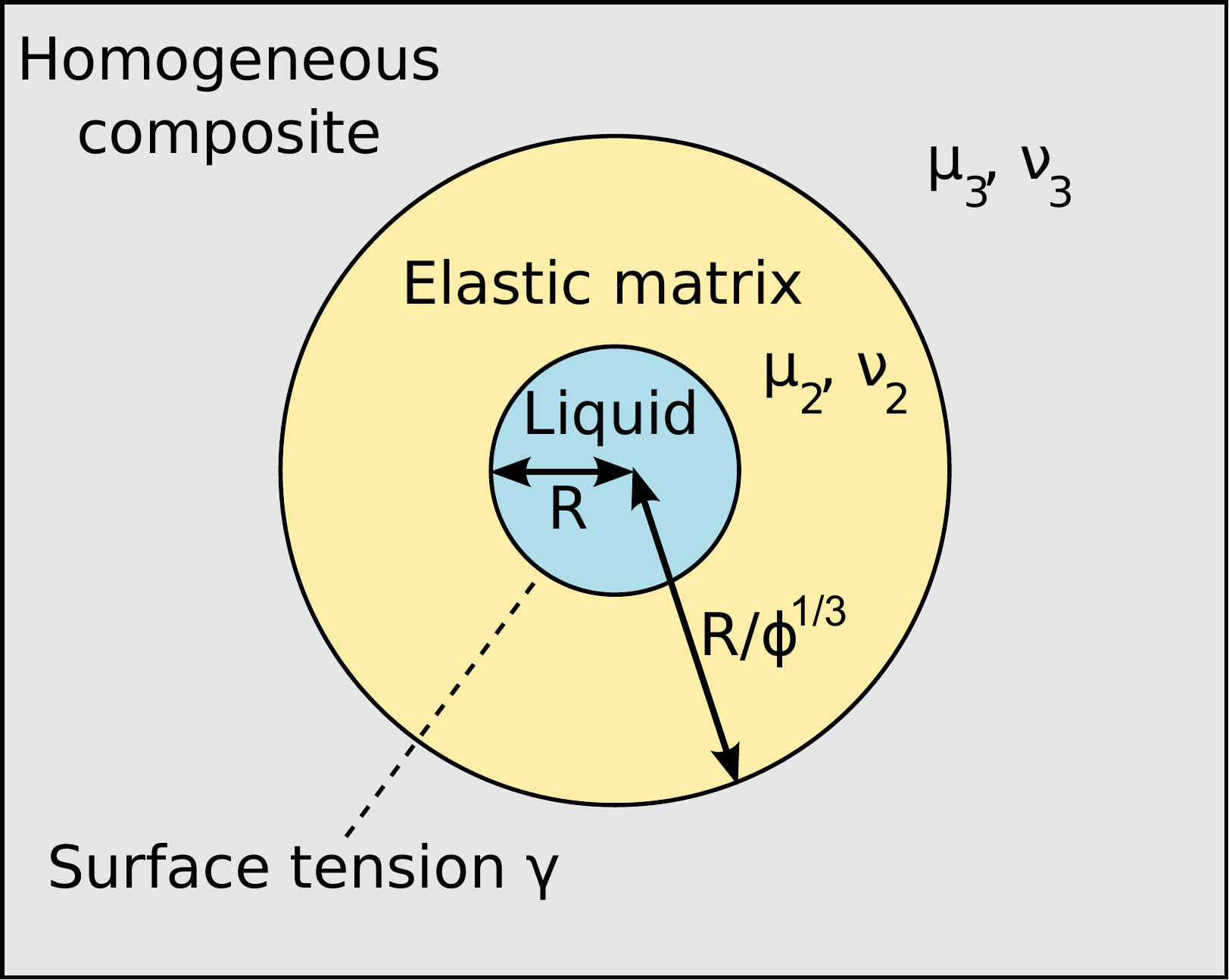}
\caption{Schematic representation of the composite material treated. Identical liquid inclusion droplets are embedded in a solid elastic matrix (subscript 2). The surrounding composite (subscript 3) is treated as an isotropic elastic medium with properties to be determined. External body forces are ignored.}
\label{figcomposite}
\end{figure}

Consider a composite system of many identical incompressible droplets embedded in an isotropic homogeneous elastic solid with shear modulus $\mu_2$ and Poisson ratio $\nu_2$ as shown in Fig. \ref{figcomposite}.
We ask how surface tension at the droplet/matrix interface affects Christensen and Lo's \cite{Christensen79} solution for the effective modulus of the composite for non-dilute droplet volume fractions $\phi$.
Our approach is based on the three-phase self-consistent model of Kerner \cite{Kerner56}.  

As noted above, the GSC approach treats the  multi-droplet system as a single \textit{composite sphere} embedded in an infinite medium of unknown effective elastic moduli $\mu_3, \nu_3$. The composite sphere consists of a liquid droplet of radius $R$, surrounded by a concentric spherical shell of matrix material of radius $R/\phi^{1/3}$, thereby preserving the liquid volume fraction $\phi$ of the original multi-droplet system. The overall approach follows that of previous work \cite[e.g.][]{Duan05JMPS, Stylesoft15}, up to a point. To make this paper reasonably self-contained we summarize the key intermediate results, and provide more detail where we believe clarity is required. 

Placing the origin of an ($r, \theta, \phi$) spherical coordinate system at the center of the composite sphere, we choose the following far-field ($r \rightarrow \infty$) displacements:
\be 
u_r^0=2\varepsilon_{A}^0 r\, \mathcal{P}_2(\cos \theta), \quad  
u_\theta^0=\varepsilon_{A}^0 r\, \frac{d \mathcal{P}_2(\cos \theta)}{d \theta}, \quad 
u_\phi^0=0\;, 
\label{remote displacements}
\ee 
where $\mathcal{P}_2$ is the Legendre polynomial of order 2; the corresponding, purely deviatoric far-field strains are:
\be
\varepsilon_{xx}^0= \varepsilon_{yy}^0=-\varepsilon_{A}^0, \quad\quad \varepsilon_{zz}^0=2 \varepsilon_{A}^0\;. 
\label{deviatoricstrain}
\ee
For the strained system, the symmetry about the $z$-axis allows the use of the following ansatz \cite[e.g.,][]{Lur'e64, Duan05,Stylesoft15} for the displacements $u_r^{(i)}$ and $u_\theta^{(i)}$ in the radial and polar directions, 
\begin{multline}
u_r^{(i)}(\rho,\theta)=\left(\mathcal{F}_i+\frac{\mathcal{G}_i}{\rho^3}\right)r+\mathcal{P}_2(\cos \theta)\times \\
\left[12 \nu_i \mathcal{A}_i \rho^2+2\mathcal{B}_i+2\frac{(5-4\nu_i)\mathcal{C}_i}{\rho^3}-3\frac{\mathcal{D}_i}{\rho^5}\right]r\;,
\label{radialansatz}
\end{multline}
\begin{multline}
u_\theta^{(i)}(\rho,\theta)=\frac{d\mathcal{P}_2(\cos \theta)}{d\theta}\times \\ \left[(7-4\nu_i)\mathcal{A}_i \rho^2+\mathcal{B}_i+2\frac{(1-2\nu_i)\mathcal{C}_i}{\rho^3}+\frac{\mathcal{D}_i}{\rho^5}\right]r\;,
\label{polaransatz}
\end{multline}
where $\rho\equiv r/R$, the index "$i$" refers to either the matrix ($i=2$) or the composite effective medium ($i=3$) phase, and $\mathcal{A}_i$ through $\mathcal{G}_i$ will be determined from the boundary conditions. 

The corresponding stress components in regions $i=~2, 3$ are
\begin{multline}
\sigma_{rr}^{(i)}(\rho,\theta)=2\mu_i 
\left\{-2 \frac{\mathcal{G}_i}{\rho^3}+\frac{\mathcal{F}_i (1+\nu_i)}{1-2\nu_i}+\right.\\
\left.\left[-6\nu_i\mathcal{A}_i \rho^2+2\mathcal{B}_i-\frac{4(5-\nu_i)}{\rho^3}\mathcal{C}_i+\frac{12 \mathcal{D}_i}{\rho^5} \right]\mathcal{P}_2(\cos \theta)\right\}\,,
\end{multline}
and
\begin{multline}
\sigma_{r\theta}^{(i)}(\rho,\theta)=2\mu_i \frac{d \mathcal{P}_2(\cos \theta)}{d\theta}\times \\
\left[(7+2\nu_i)\mathcal{A}_i \rho^2+\mathcal{B}_i +\frac{2(1+\nu_i)}{\rho^3}\mathcal{C}_i  -\frac{4 \mathcal{D}_i}{\rho^5} \right]\,.
\end{multline}
The relation between the pressure $p$ and the components of the hydrostatic stress tensor in the liquid region (i=1) is:
\be 
\sigma_{rr}^{(1)}=\sigma_{\theta\theta}^{(1)}=-p, \quad\quad \sigma_{r\theta}^{(1)}=0\,.
\ee

Now, combining the stress-displacement relationship with Eq. (\ref{remote displacements}) gives the far-field stresses:
\be
\sigma_{rr}^0=4 \varepsilon_A^0 \mu_3 \mathcal{P}_2(\cos \theta)\,,   \quad 
\sigma_{r\theta}^0=2 \varepsilon_A^0 \mu_3 \frac{d \mathcal{P}_2(\cos \theta)}{d \theta}\,.
\ee
Three constants are determined by the far-field stress/strain, viz.,  $\mathcal{A}_3=0$, $\mathcal{B}_3=\varepsilon_A^0$, and $\mathcal{F}_3=0$.  There are ten equations for the remaining ten unknowns; $p, \mathcal{A}_2, \mathcal{B}_2, \mathcal{C}_2, \mathcal{D}_2, \mathcal{F}_2, \mathcal{G}_2, \mathcal{C}_3, \mathcal{D}_3, \mathcal{G}_3$.  Six equations arise from continuity of strain and stress at the composite sphere surface ($\rho=\alpha \equiv 1/ \phi^{1/3}$), one from the incompressibility of the droplet and three from the stress boundary conditions associated with the generalized Young-Laplace condition.  Taking these in turn, we have
\begin{align}
u_r^{(2)}(\alpha, \theta)&=u_r^{(3)}(\alpha, \theta),\quad u_\theta^{(2)}(\alpha, \theta)=u_\theta^{(3)}(\alpha, \theta) \qquad \textrm{and} \nonumber \\
\sigma_{rr}^{(2)}(\alpha, \theta)&=\sigma_{rr}^{(3)}(\alpha, \theta),\quad \sigma_{r\theta}^{(2)}(\alpha, \theta)=\sigma_{r\theta}^{(3)}(\alpha, \theta), 
\label{continuityexternalur}
\end{align}
where the continuity of strain for $u_r$ ($u_\theta$) provides two (one) equations and that for stress $\sigma_{rr}$ ($\sigma_{r\theta}$) provides two (one) equations; droplet incompressibility requires that
\be
\int\limits_{S^{int}}  u_r^{(2)}(1,\theta) R^2 \sin \theta\, d\theta\, d\phi=R^3(\mathcal{F}_2+ \mathcal{G}_2)=0, \quad
\ee
where $S^{int}$ denotes the droplet/matrix interface. The final three equations arise from the stress boundary conditions at the surface of the droplet, treated using the generalized Young-Laplace condition \cite{Style15, Stylesoft15} written as 
\be
\sigma \cdot \textbf{n}=-p\textbf{n}+\gamma \mathcal{K} \textbf{n}, 
\label{YoungLaplace}
\ee
where \textbf{n} is the normal to the droplet surface and $\mathcal{K}$ is its curvature, while the surface stress is taken as an isotropic and strain-independent surface tension, $\gamma$. Using leading-order expressions for $\textbf{n}$ and $\mathcal{K}$ \cite{Stylesoft15},
%
equation (\ref{YoungLaplace}) becomes
\begin{multline}
\left[\begin{array}{c}\sigma_{rr}+\sigma_{r\theta}\,(u_\theta-\frac{\partial u_r}{\partial \theta})R^{-1} \\ \sigma_{\theta r}+\sigma_{\theta\theta}\,(u_\theta-\frac{\partial u_r}{\partial \theta})R^{-1}\end{array}\right]^{(2)}-   
\left[\begin{array}{c}-p+0 \\ 
0-p(u_\theta-\frac{\partial u_r}{\partial \theta})R^{-1}\end{array}\right]^{(1)}=\\
= \frac{\gamma}{R^2}\left[\begin{array}{c} 2R-(2 u_r+ \cot \theta \frac{\partial u_r}{\partial \theta}+\frac{\partial^2 u_r}{\partial \theta^2}) \\ 2(u_\theta-\frac{\partial u_r}{\partial \theta})\end{array}\right]\;.
\label{vectoryoung}
\end{multline}
The radial component of equation (\ref{vectoryoung}) implies that $p=2\gamma/R$. Hence, at leading order in $\textbf{u}$, the droplet pressure is \textit{unaffected} by the externally applied strain (\ref{deviatoricstrain}). This linearizes the single equation associated with the $\theta$-component, thus simplifying the overall system and the ten unknowns $p, \mathcal{A}_2, \mathcal{B}_2, \mathcal{C}_2, \mathcal{D}_2, \mathcal{F}_2, \mathcal{G}_2, \mathcal{C}_3, \mathcal{D}_3, \mathcal{G}_3$ are determined by the ten equations (\ref{continuityexternalur})-(\ref{vectoryoung}) in terms of the unknown parameters $\mu_3$, $\nu_3$.  To find $\mu_3$, we use the \emph{energetic self-consistency condition} that the total work associated with the presence of the composite sphere inside the infinite effective medium, $W=0$ \cite{Christensen79}.
%
We evaluate  $W$ in the spirit of Eshelby \cite[Eq. 5.1 of][]{Eshelby56} to find \cite{Stylesoft15} 
\begin{align}
W = & \frac{1}{2}\int_{S^{ext (+)}}[n_i \sigma_{ij}^0 u_j-n_i \sigma_{ij} u_j^0]dS \nonumber \\
&-\frac{\gamma}{2}\int_{S^{int}} \mathcal{K} u\cdot n\, dS+\gamma \Delta S, 
\label{workcompositeinclusion}
\end{align}
where $S^{ext (+)}$ denotes the external side of the composite sphere/external effective medium interface, and we note that the last two summands exactly cancel (this was not pointed out in previous work \cite{Stylesoft15}). 

The condition $W=0$ provides a constraint in the form of a quadratic equation, which determines the relative effective shear modulus $\mu_3/\mu_2$. Indeed, up to second order in $\textbf{u}$, we can replace the normal vector $\textbf{n}$ in (\ref{workcompositeinclusion}) by the basis unit vector $\hat{r}$ of the spherical coordinate system and find that 
\begin{align}
W=&\frac{1}{2}\int_{S^{ext (+)}}[\sigma_{rr}^0 u_r+\sigma_{r\theta}^0 u_\theta - \sigma_{rr} u_r^0 - \sigma_{r\theta} u_\theta^0]dS \nonumber \\
&=-\frac{48 \pi \mu_3 R^3 \varepsilon_A^0 (\nu_3-1)}{\alpha^2}\, \mathcal{C}_3\;.
\end{align}
Therefore, $\mathcal{C}_3=0$, and 
plugging this into the solution of the system of equations (\ref{continuityexternalur})-(\ref{vectoryoung}) yields a quadratic condition for the relative effective shear modulus $\mu_{rel}\equiv \mu_3/\mu_2$  as a function of $\phi$, $\nu_2$, and ${\gamma}/{ (R \mu_2)}$ as follows
\be
2R\mu_2(a_0+a_1 \mu_{rel}+a_2 \mu_{rel}^2)+ \gamma (b_0+b_1 \mu_{rel}+b_2 \mu_{rel}^2)=0\,,
\label{quadraticgeneral}
\ee
where the coefficients are in Appendix \ref{AppendixA}.  

\begin{figure}[]
\centering 
\includegraphics[width=1\columnwidth]{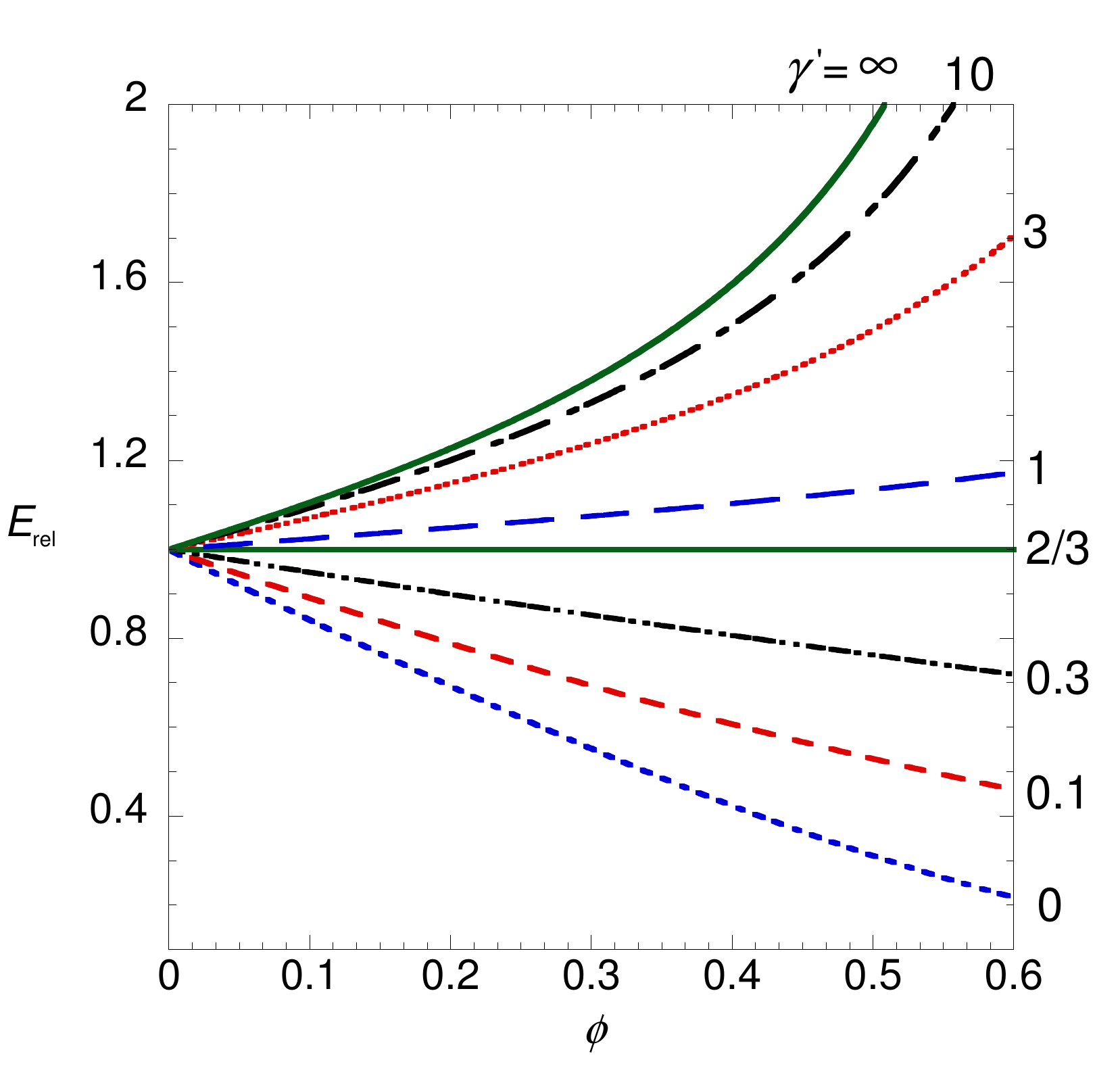}
\caption{$E_{rel}$ against $\phi$ over a wide range of $\gamma^\prime$ (as labeled in the figure) from the softening to the stiffening regime for the incompressible case $\nu_2=1/2$.}
\vspace{-0.4cm}
\label{fig:regular}
\end{figure}

\begin{figure}[htp]
\centering 
\includegraphics[width=1\columnwidth]{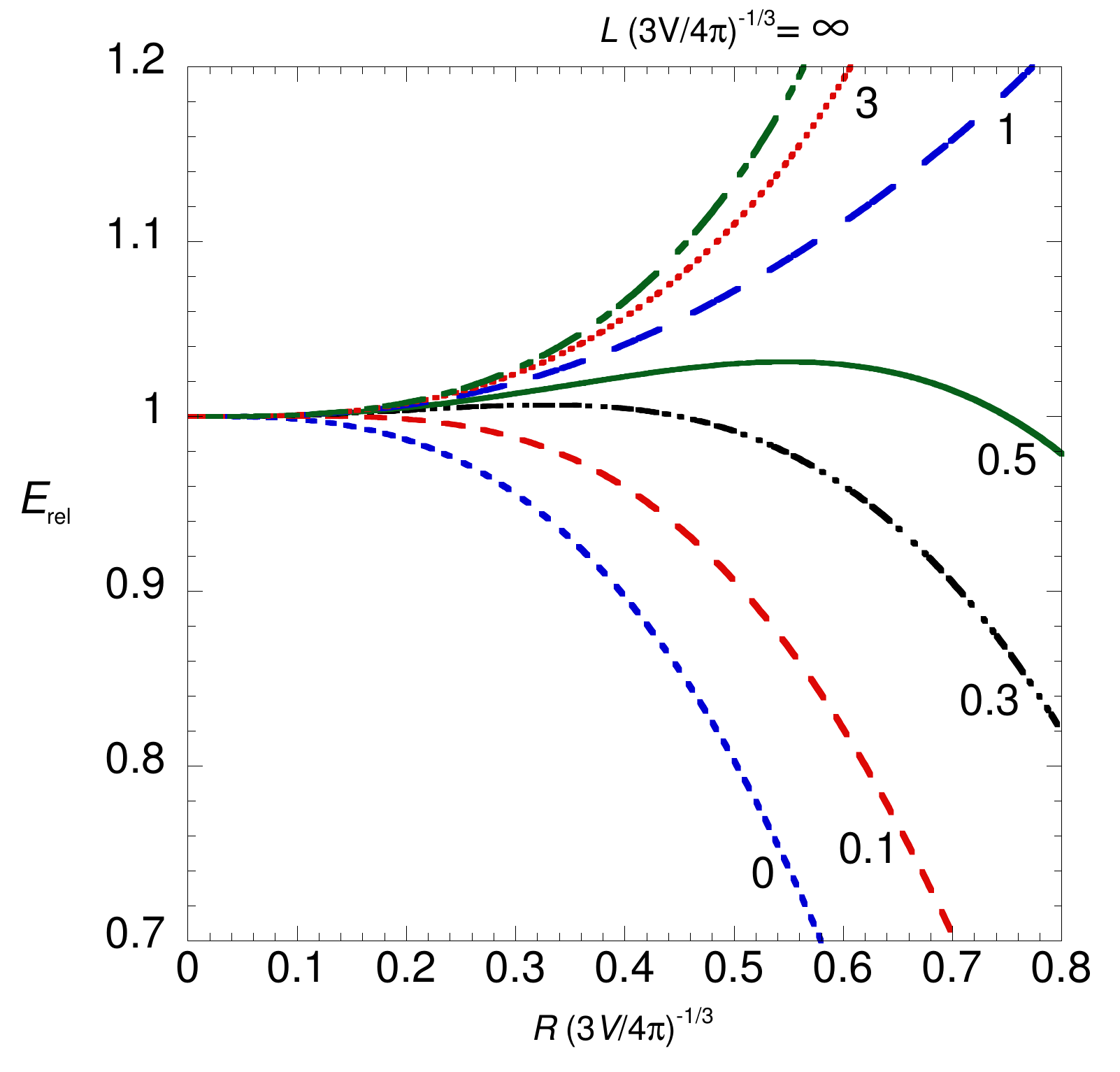}
\vspace{-0.4cm}
\caption{$E_{rel}$ against $R/[ (3V / 4 \pi )^{1/3}]$ for a wide range of the parameter $L/[ (3V / 4 \pi )^{1/3}]$ (as labeled in the figure) for the incompressible case $\nu_2=1/2$}
\label{fig:L}
\end{figure}

For the remainder of the paper we will focus on the special case of an incompressible matrix, for which $E_{rel}\equiv\left(\frac{E_3}{E_2}\right)=\left(\frac{\mu_3}{\mu_2}\right)\equiv\mu_{rel}$ and $\nu_2=1/2$. Considering the elastocapillarity length $L\equiv \gamma/E_2$ based on the matrix phase of the composite sphere, we define the dimensionless parameter $\gamma^\prime\equiv L/R=\gamma/(E_2 R)$. 
Fig. \ref{fig:regular} shows the behavior of $E_{rel}$ as a function of $\phi$, and in Fig. \ref{fig:L} $E_{rel}$ is plotted against $R/[ (3V / 4 \pi )^{1/3}]$, where $V$ is the outer sphere volume in the GSC framework (i.e., $E_{rel}$ is plotted against $\phi^{1/3}$).  Clearly, Fig. \ref{fig:regular} shows monotonic response over a large range of $\phi$, exhibiting  softening (stiffening) $\gamma^\prime < 2/3$ ($\gamma^\prime > 2/3$) behavior that spans the experimental range seen by Style et al., \cite{Style15}.  Moreover, the dilute theory \cite{Stylesoft15} is quantitatively captured in the limit $\phi\rightarrow 0$ of the present theory.
Furthermore, we find exact ``mechanical cloaking'', in which $E_{rel}$ is constant at $\gamma^\prime = 2/3$ for all liquid volume fractions.  Exactly the same cloaking condition is found in the dilute theory \cite{Stylesoft15}, {\em and} from a complimentary Mori-Tanaka approach \cite{MSWa} as described in Fig. \ref{figcomparison} below.

In the stiffening regime, as droplets become small and $\gamma^\prime$ becomes large, the quadratic condition (\ref{quadraticgeneral}) for $\mu_{rel}$ simplifies to $b_0+b_1 \mu_{rel}+b_2 \mu_{rel}^2=0$.
At a given $\phi$ the solution of this equation, $\mu_{rel}=\mu_{rel,R \ll L}[\phi,\nu_2]$, gives the upper limit of rigidity among all $\gamma^\prime$-curves, showing a stiffening behavior proportional to $\frac{1}{1-\phi}$ in the limit $\phi\rightarrow 0$  ($\gamma^\prime=\infty$-line in Fig. \ref{fig:regular}).    

Now we examine the deformation of the inclusion phase by making use of Eqs. (\ref{dropshaper}) and (\ref{dropshapetheta}) in Appendix \ref{AppendixB} to evaluate the effective droplet strain $\varepsilon_d\equiv(l-2R)/R=2 u_r(1,0)/R$, where $l$ denotes the major axis of the droplet\footnote{Note that we could well have chosen $\varepsilon_d=(l-2R)/2R$ rather than $(l-2R)/R$, but chose the latter to facilitate comparison to the dilute results from \cite{Stylesoft15}.}.  In terms of $\alpha=\phi^{-1/3}$, $\gamma^\prime$, and the solution of Eq.(\ref{quadraticgeneral}), $\mu_{rel}=\mu_3/\mu_2$, the radial and polar displacements of the droplet interface are
\be
\frac{u_r(1,\theta)}{R}=100 \mu_{rel} \alpha^3\,\varepsilon_A^0 \; \frac{f_1}{f_2+\gamma^\prime f_3}\;\mathcal{P}_2(\cos \theta)
\label{radshape}
\ee
and
\be 
\frac{u_\theta(1,\theta)}{R}=25 \mu_{rel} \alpha^3\,\varepsilon_A^0 \;\frac{f_4+\gamma^\prime f_5}{f_2+\gamma^\prime f_3}\;\frac{d \mathcal{P}_2(\cos \theta)}{d \theta}\,,
\label{thetashape}
\ee
where the coefficients $f_1$--$f_5$ are in Appendix \ref{AppendixB}.
When $R \ll L$ and $\gamma^\prime \gg 1$, and the radial displacement becomes extremely small, then
the inclusions remain spherical. In the opposite limit, $R \gg L$ and $\gamma^\prime \ll 1$,
the inclusion shape is again scale-invariant, in agreement with the theory for the case of pure bulk elasticity.
The corresponding effective droplet strain is
$\varepsilon_d = 200 \mu_{rel} \alpha^3 \varepsilon_A^0 f_1/(f_2+\gamma^\prime f_3)$.
In the dilute limit $\phi\rightarrow 0$ of the incompressible case, the droplet's effective strain and shape reduce to 
$ \varepsilon_d\vert_{\phi\rightarrow 0}=40 \varepsilon_A^0/(6+15 \gamma^\prime)$,
\be
\nonumber
\frac{u_r(1,\theta)}{R}\vert_{\phi\rightarrow 0}=\frac{5}{3} \varepsilon_A^0 \frac{1+3 \cos(2 \theta)}{2+5 \gamma^\prime}\,,
\ee
and
\be
\nonumber
\frac{u_\theta(1,\theta)}{R}\vert_{\phi\rightarrow 0}=-\frac{5}{2} \varepsilon_A^0 \frac{(2+3\gamma^\prime) \sin(2 \theta)}{2+5 \gamma^\prime} \;, 
\ee
thereby recovering the results of Style et al. \cite{Stylesoft15}.   Interestingly, we find that in this limit, at the exact cloaking point ($\gamma^\prime=2/3$), the droplet will stretch less than the host material viz., $(l-2R)/(2R)=(5/8) \varepsilon_{zz}^0$.  However, the droplet will stretch the same as the host material at $\gamma^\prime= 4/15 < 2/3$, within the softening regime. The predictions of Eshelby's theory \cite{Eshelby57} and effective droplet strain $(10/3) \varepsilon_{zz}^\infty$ are recovered for $\phi,\gamma^\prime \ll 1$, whereas an unperturbed spherical shape is found when $\gamma^\prime \gg 1$, and arbitrary values of $\phi$. 

Finally, in Fig. \ref{figcomparison}, we compare $E_{rel}$ for this theory (red) with a modified version of the Mori-Tanaka theory (green \cite{MSWa}) and the dilute theory (blue \cite{Stylesoft15}).  We see that this theory predicts a more pronounced softening in the softening regime ($\gamma^\prime < 2/3$), and a more pronounced stiffening in the stiffening regime ($\gamma^\prime > 2/3$) than does the modified Mori-Tanaka theory.  Interestingly, in the stiffening regime, the three-phase and dilute theories are perhaps experimentally indistinguishable, well beyond the concentration range where the latter breaks down ($\phi \approx 0.2$).  This indicates that, depending on the range of $\gamma^\prime$ of relevance, the dilute theory provides a simple framework for comparison with experiment given that it is the appropriate asymptotic limit of the non-dilute theory.  All three theories predict exactly the same mechanical cloaking condition, $\gamma^\prime=2/3$, of the inclusions, independent of $\phi$.  We note here that the results in the surface-tension free limit $\gamma \rightarrow 0$ are compared with the classical result \cite{Christensen79} in Appendix \ref{AppendixA}.

\begin{figure*}[t]
\centering 
\begin{tabular}{ccc}
\includegraphics[width=0.68\columnwidth]{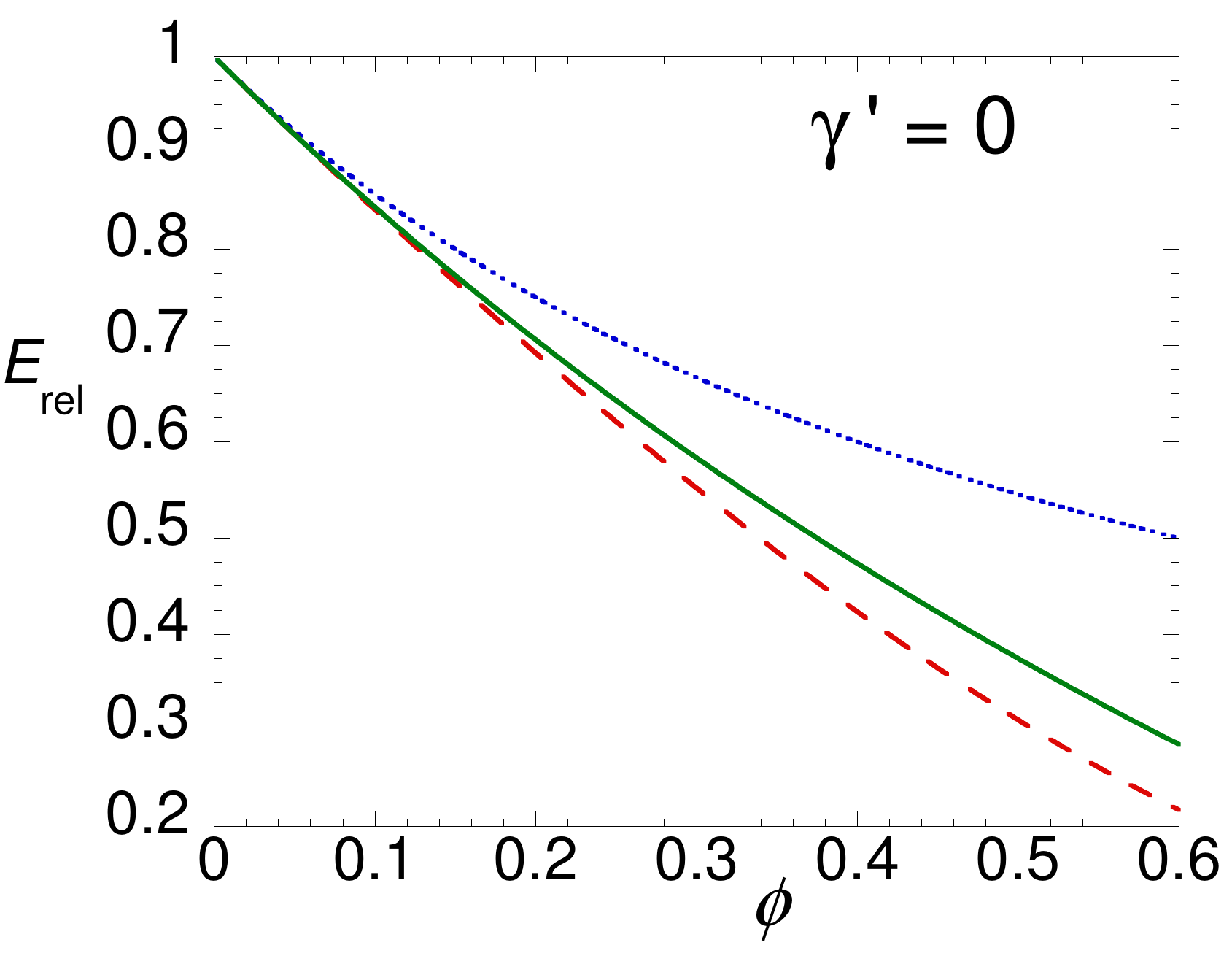}&
\includegraphics[width=0.68\columnwidth]{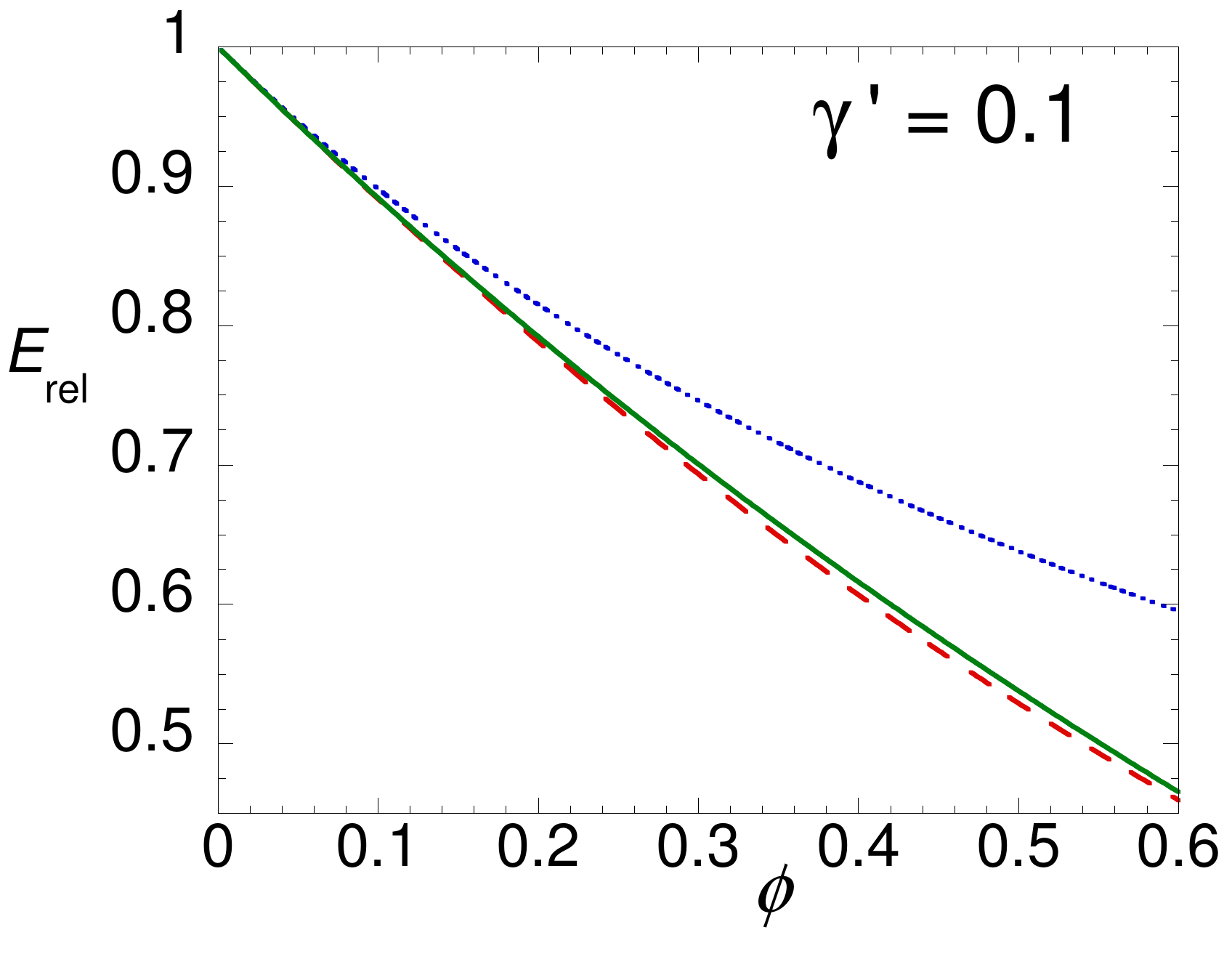}&
\includegraphics[width=0.68\columnwidth]{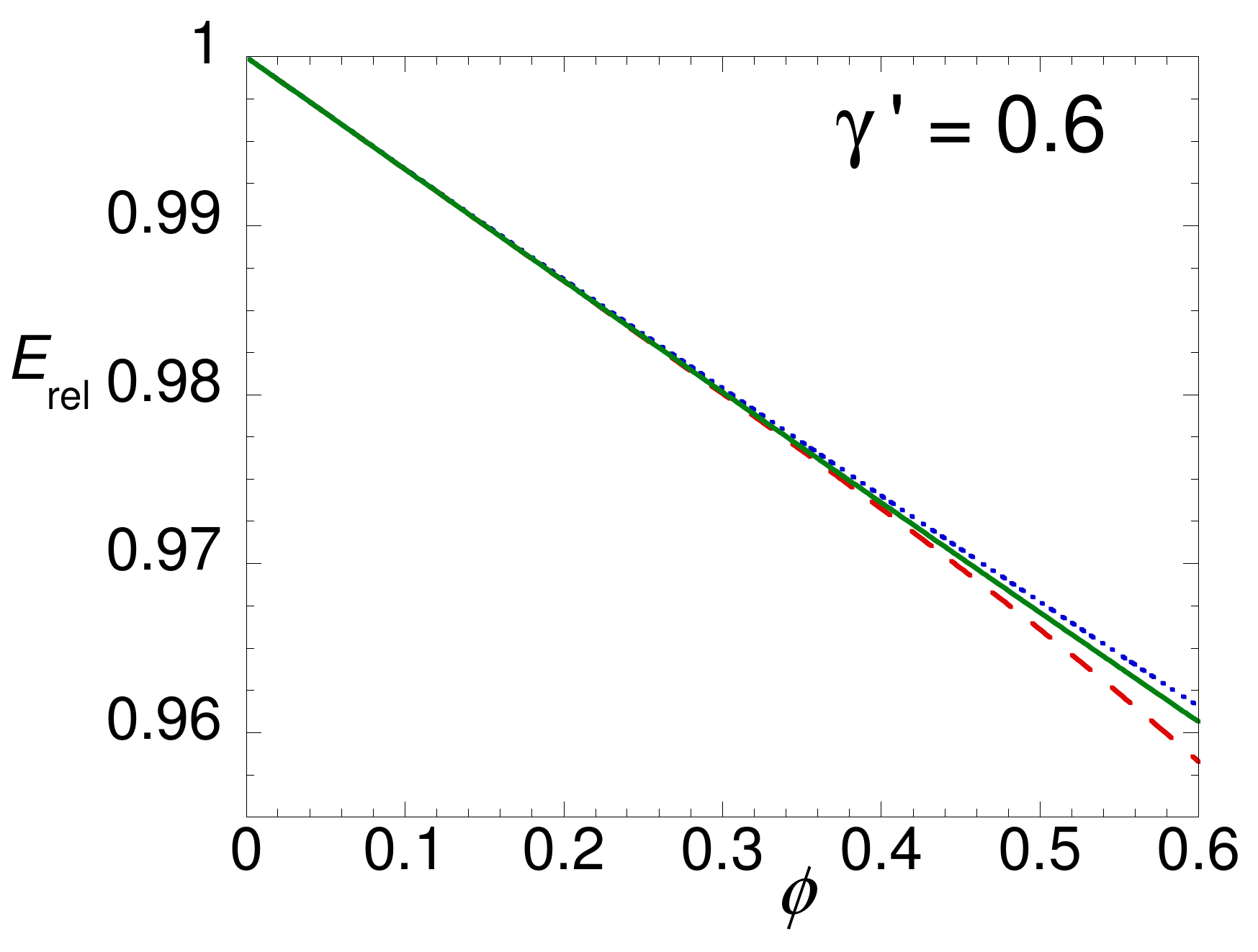}\\
\includegraphics[width=0.68\columnwidth]{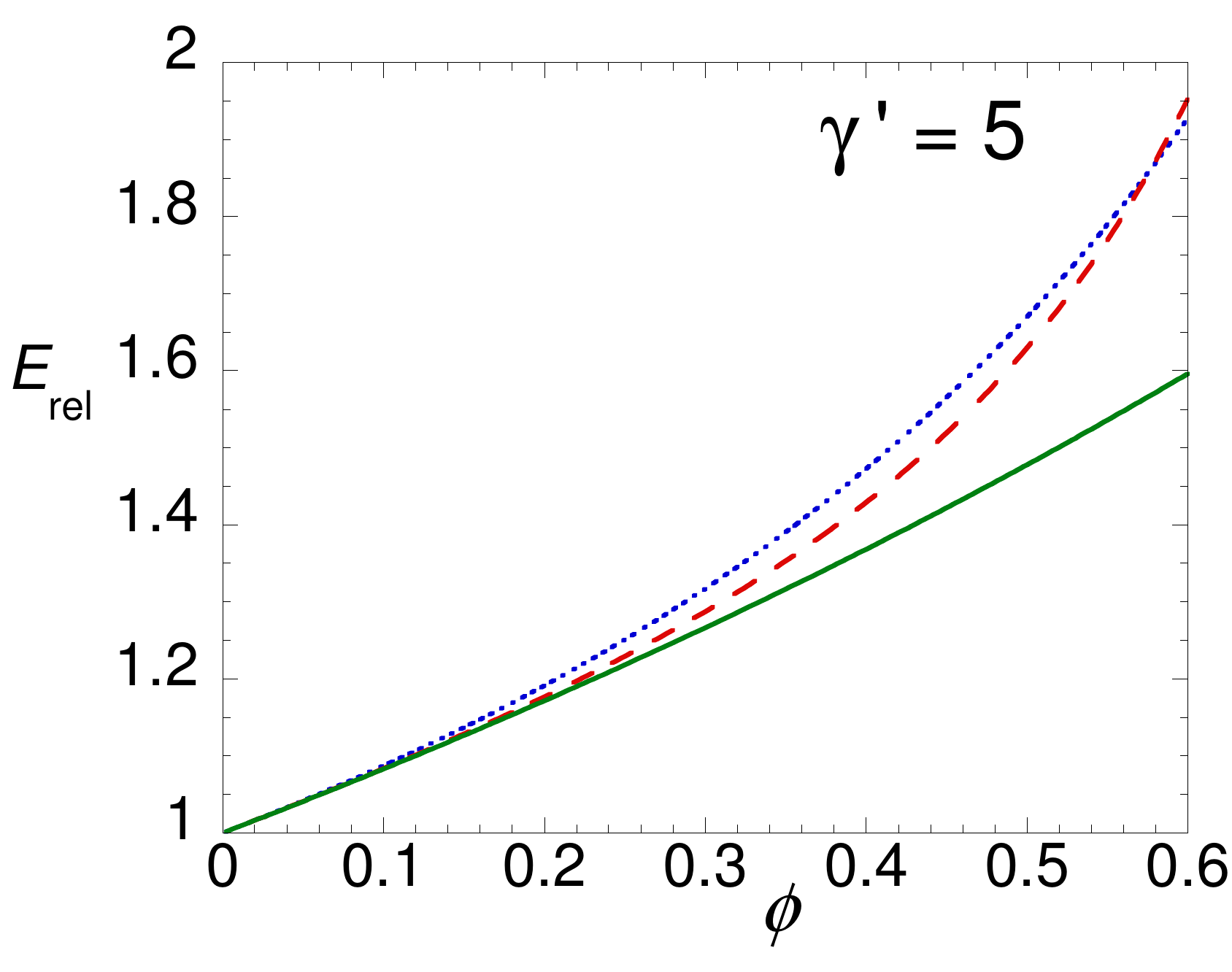}&
\includegraphics[width=0.68\columnwidth]{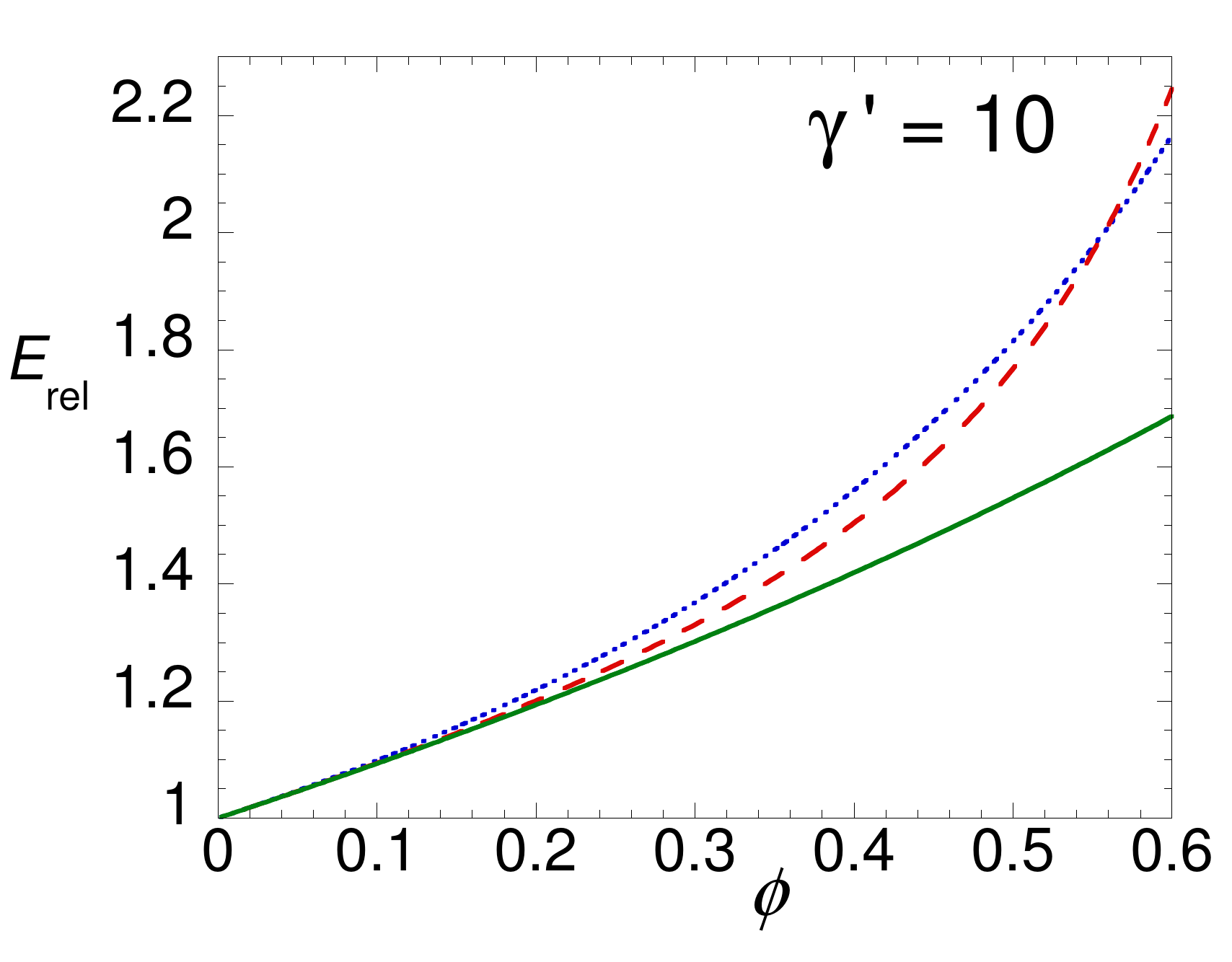}&
\includegraphics[width=0.68\columnwidth]{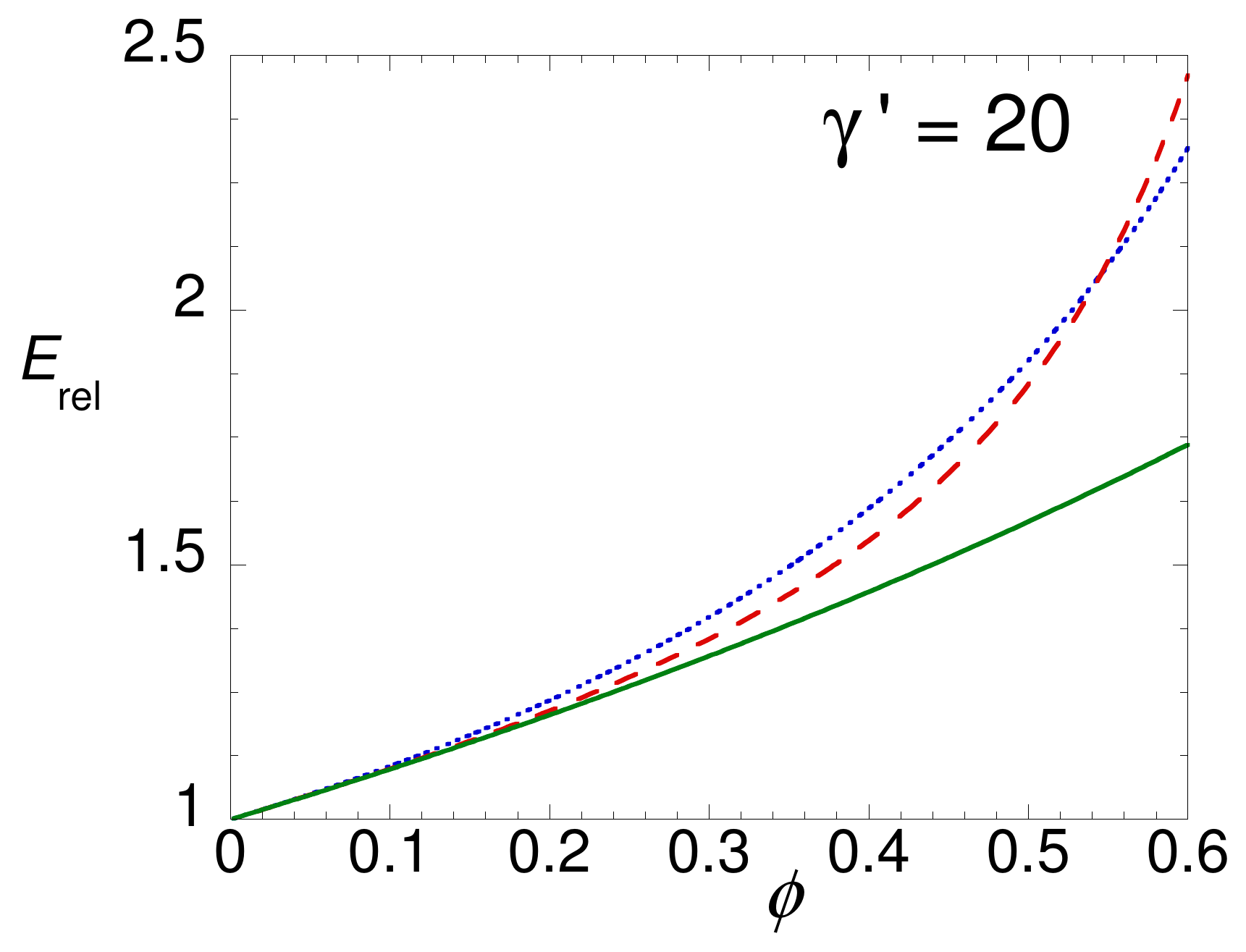}
\end{tabular}
\caption{$E_{rel}$ versus $\phi$ in the incompressible matrix case for the dilute (blue dotted, \cite{Stylesoft15}), modified Mori-Tanaka (green continuous,  \cite{MSWa}), and the present three-phase GSC (red dashed) theories for a range of $\gamma^\prime$ from the softening to the stiffening regime. In the top row $\gamma^\prime=0$, $0.1$, and $0.6$ and in the bottom row $\gamma^\prime= 5$, $10$, and $20$. Note the vertical axes have different scales to show the behavior with $\phi$ across the range of $\gamma^\prime$ shown. The softening ($\gamma^\prime < 2/3$) and stiffening ($\gamma^\prime > 2/3$) exhibited by the present three-phase GSC theory are more pronounced than in the modified Mori-Tanaka theory.  Clearly, the dilute theory, when extended beyond its range of validity, shows a stiffening effect quantitatively similar to the three-phase GSC theory, whereas in softening the latter theory more closely coincides with the modified Mori-Tanaka theory.  All three approaches predict exact mechanical cloaking of the inclusions (i.e. $E_{rel}=1$) at $\gamma^\prime= 2/3$.}
\label{figcomparison} 
\end{figure*} 

\section{Conclusions}

Based on a three-phase generalized self-consistent approach, we have estimated elastic moduli of composites including liquid droplets, by taking into account the (linear-elastic) solid/droplet interfacial surface tension. In the limit $\phi \rightarrow 0$, we recover the dilute-theory expressions of Style et al., \cite{Style15, Stylesoft15}.   The Young's modulus of the composite depends on $\gamma^\prime$, which is the ratio of the elastocapillary length $L$, to the inclusion radius $R$.  The results are compared quantitatively with the dilute theory and a version of Mori-Tanaka theory, both of which include surface tension.  In the softening case, the three-phase theory and the modified Mori-Tanaka theory are consistent over a wide range of $\gamma^\prime$, whereas in the stiffening case the three-phase theory is consistent with the dilute theory even for volume fractions over which the latter is expected to break down.  All three models predict cloaking of the far-field effects associated with the inclusions when $R=3L/2$ or $\gamma^\prime = 2/3$ for all volume fractions $\phi$.  We have calculated both the effective droplet strain and radial and polar interfacial displacements and found that, in the dilute limit and at exact cloaking, the droplet strain is smaller than that of the host material, whereas they are equal at $\gamma^\prime=4/15 < 2/3$, within the softening regime. 

Finally, we note that there is an interesting similarity between the mechanical response of the multi-phase soft materials studied here and what one finds in poroelasticity, which is a framework used to study the effective medium response of fluid filled host structures, applied to 
 problems ranging from biology to geophysics \cite[e.g.,][]{Forterre, Moeendarbary, Hesse, MacMinn:2012, Wang, MacMinn:2015}.  For example, in many biological settings, the composite medium has soft elastic or liquid inclusions, and the deformation of the host material is controlled by the value of $\phi$, which is typically determined as part of the solution to the problem.  Whereas in poroelasticity a major challenge involves the modeling of the flow permeability, which is {\em specified} as a function of $\phi$, our approach {\em derives} the mechanical response as a function of $\phi$.  We suggest that by treating the mechanical properties of poroelastic media within the framework studied here, one can constrain the $\phi$ dependence of transport properties such as the flow permeability.


\section{Acknowledgments}\label{Acknowledgments}
FM and JSW acknowledge Swedish Research Council Grant No. 638-2013-9243 and the 2015 Geophysical Fluid Dynamics Summer Study Program at the Woods Hole Oceanographic Institution, which is supported by the National Science Foundation and the Office of Naval Research under OCE-1332750.  JSW also acknowledges 
a Royal Society Wolfson Research Merit Award.  

\eject
\appendix

\section{Coefficients in Equation (\ref{quadraticgeneral})  \label{AppendixA}}

\begin{widetext}

The coefficients in equation (\ref{quadraticgeneral}) are

\be
\left\{
\begin{array}{l}
a_0=-49 - 252 \alpha^5 + 25 \nu_2^2 - 25 \alpha^3 (-7 +\nu_2^2) - 
 25 \alpha^7 (-7 + \nu_2^2) + \alpha^{10} (-49 + 25 \nu_2^2)\\
a_1=-7 + 504 \alpha^5 + 30 \nu_2 + 150 \alpha^7 (-3 + \nu_2) \nu_2 + 
 25 \nu_2^2 + 50 \alpha^3 (-7 + \nu_2^2) - 
 3 \alpha^{10} (49 - 140 \nu_2 + 75 \nu_2^2)\\
a_2=56 - 252 \alpha^5 - 30 \nu_2 - 50 \nu_2^2 - 
 25 \alpha^3 (-7 + \nu_2^2) + 
 50 \alpha^7 (7 - 12 \nu_2 + 8 \nu_2^2) + 
 4 \alpha^{10} (49 - 105 \nu_2 + 50 \nu_2^2)
\end{array}
\right., 
\ee
and 
\be
\left\{
\begin{array}{l}
b_0=252 \alpha^5 (-1 + 2 \nu_2) + 25 \alpha^7 (-7 + \nu_2^2) - 
 50 \alpha^3 (-7 + 6 \nu_2 + 4 \nu_2^2) + 
 4 (49 - 63 \nu_2 + 20 \nu_2^2) + \alpha^{10} (-119 + 48 \nu_2 + 
    95 \nu_2^2)\\
b_1=-150 \alpha^7 (-3 + \nu_2) \nu_2 - 504 \alpha^5 (-1 + 2 \nu_2) + 
 100 \alpha^3 (-7 + 6 \nu_2 + 4 \nu_2^2) + 
 4 (7 - 39 \nu_2 + 20 \nu_2^2) - 
 3 \alpha^{10} (119 - 388 \nu_2 + 285 \nu_2^2)\\
b_2=2(126 \alpha^5 (-1 + 2 \nu_2) - 
  25 \alpha^3 (-7 + 6 \nu_2 + 4 \nu_2^2) - 
  25 \alpha^7 (7 - 12 \nu_2 + 8 \nu_2^2) - 
  4 (28 - 51 \nu_2 + 20 \nu_2^2) + \alpha^{10} (238 - 606 \nu_2 + 
     380 \nu_2^2))
\end{array}
\right.. 
\ee

We note here that Eq. (3.14) of Christensen and Lo \cite{Christensen79} is incorrect in the regime of large droplets ($R \gg L$). In this regime, the condition (\ref{quadraticgeneral}) reduces to
$a_0+a_1 \mu_{rel}+a_2 \mu_{rel}^2=0$,
whose solution $\mu_{rel}=\mu_{rel,R\gg L}[\phi,\nu_2]$ is invariant under $\mu_2$-scalings, as expected from the symmetries of the equations of the elastostatics.
\end{widetext}

\section{Shape of the droplets under uniaxial stress \label{AppendixB}}

By using our model, we determine the shape of the generic droplet embedded in an incompressible matrix ($\nu_2=1/2$) undergoing uniaxial stress. Note that, with $\varepsilon_A^0=\varepsilon_{zz}^0/2$, the purely deviatoric far-field strain conditions of Eqs. (\ref{deviatoricstrain}) are equivalent to the strain system of Style et al., \cite[][see 3 lines below Eq.(2)]{Stylesoft15}.  Substituting the expressions for $\mathcal{A}_2$ - $\mathcal{G}_2$ into Eqs. (\ref{radialansatz}) and (\ref{polaransatz}), we obtain the following surface displacements, 
\be 
\frac{u_r(1,\theta)}{R}=[6\mathcal{A}_2+2\mathcal{B}_2+6\mathcal{C}_2-3\mathcal{D}_2]\;\mathcal{P}_2(\cos \theta)\;,
\label{dropshaper}
\ee
and
\be
\frac{u_\theta(1,\theta)}{R}=[5\mathcal{A}_2+\mathcal{B}_2+\mathcal{D}_2]\;\frac{d \mathcal{P}_2(\cos \theta)}{d \theta}\;.
\label{dropshapetheta}
\ee
Finally, we note that the coefficients $\{f_i\}$ in Eqs. (\ref{radshape}) and (\ref{thetashape}) are: 
\begin{widetext}
\be \left\{\begin{array}{l}
f_1=40 (\mu_{rel}-1) - 21 (\mu_{rel}-1) \alpha^2 + (19 + 16 \mu_{rel}) \alpha^7 \\
f_2=6 (38 (\mu_{rel}-1)^2 + 75 (\mu_{rel}-1) (2 + 3 \mu_{rel}) \alpha^3 + 
   112 (2 + \mu_{rel} - 3 \mu_{rel}^2) \alpha^5 + 
   50 (\mu_{rel}-1) (3 + 4 \mu_{rel}) \alpha^7 + (2 + 3 \mu_{rel}) (19 + 
      16 \mu_{rel}) \alpha^{10}) \\
f_3= 15 (-48 (\mu_{rel}-1)^2 + 40 (\mu_{rel}-1) (2 + 3 \mu_{rel}) \alpha^3 + 
   30 (3 + \mu_{rel} - 4 \mu_{rel}^2) \alpha^7 + (2 + 3 \mu_{rel}) (19 + 16 \mu_{rel}) \alpha^{10})\\
f_4=10 (\mu_{rel}-1) + 28 (\mu_{rel}-1) \alpha^2 + 2 (19 + 16 \mu_{rel}) \alpha^7 \\
f_5=3 (40 (\mu_{rel}-1) - 56 (\mu_{rel}-1) \alpha^2 + (19 + 16 \mu_{rel}) \alpha^7)\quad.
\end{array}\right.
\ee
\end{widetext}



\begin{thebibliography}{43}

\bibitem{Alain:2015} L. A. Mihai, K. Alayyash and A. Goriely, 
\textit{Proceedings of the Royal Society of London A}, 2015, \textbf{471}, 20150107. 
\bibitem{Bertoldi:2015} C. Coulais, J. T. B. Overvelde, L. A. Lubbers , K. Bertoldi and M. van Hecke, 
\textit{Phys. Rev. Lett.}, 2015, \textbf{115}, 044301. 
\bibitem{Hashin64} Z. Hashin and W. Rosen, \textit{J. Appl.  Mech.}, 1964, \textbf{31}, 223-232.
\bibitem{Hashin62} Z. Hashin, 
\textit{J. Appl. Mech., Trans. ASME}, 1962, \textbf{29}, 143-150.
\bibitem{Kroner58} E. Kr\"oner, 
\textit{Zeitschrift f\"ur Physik}, 1958, \textbf{151}, 504-518. 
\bibitem{Budiansky65} B. Budiansky, 
\textit{J. Mech. Phys. Solids}, 1965, \textbf{13}, 223-227.
\bibitem{Hill65a}  R. Hill, 
\textit{J. Mech. Phys. Solids}, 1965, \textbf{13}, 213-222.
\bibitem{Hill65b} R. Hill, 
\textit{J. Mech. Phys. Solids}, 1965, \textbf{13}, 189-198.   
\bibitem{Kerner56} E. H. Kerner, 
\textit{Proc. Phys. Soc. B}, 1956, \textbf{69}, 808-813.
\bibitem{footnote1} see also e.g. ref. \onlinecite{Christensen79} or refs. \onlinecite{Smith74,Smith75}, although unjustified assumptions are invoked in these two latter papers.  Note further that the relative effective shear modulus formula $\mu/\mu_m$ presented  in ref. \onlinecite{Christensen79} (Eqs. 3.14 through 3.18) should be revised.  This is evident from its symmetry breaking property under arbitrary rescaling of the matrix shear modulus $\mu_m$, whereas $\mu/\mu_m$ is expected to be invariant in the case of (interface stress-free) liquid inclusions. Such a symmetry breaking is clear as $\eta_2$ in their Eq. (3.18), and hence also the solution $\mu/\mu_m$ of the quadratic Eq. (3.14), is $\mu_m$-~dependent even in the liquid inclusion limit $\mu_i \rightarrow 0 $.
\bibitem{Pol58} C. van der Poel, 
\textit{Rheol. Acta}, 1958, \textbf{1}, 198-205.
\bibitem{Christensen79} R. M. Christensen and K. H. Lo, 
\textit{J. Mech. Phys. Solids}, 1979, \textbf{27}, 315-330.
\bibitem{Shick94} R. A. Shick and H. Ishida,
in \textit{Characterization of Composite Materials}, ed. Hatsuo Ishida, Momentum Press LLC, New York, reprint edition, 2010, Chp.8, pp.148-183.
\bibitem{Takahashi78} K. Takahashi, M. Ikeda, K. Harakawa, K. Tanaka and T. Sakai, 
\textit{J. Polym. Sci.: Polym. Phys.}, 1978, \textbf{16}, 415-425.
\bibitem{Mora10} S. Mora, T. Phou, J.-M. Fromental, L. M. Pismen and
Y. Pomeau, \textit{Phys. Rev. Lett.}, 2010, \textbf{105}, 214301.
\bibitem{Mora11} S. Mora, M. Abkarian, H. Tabuteau and Y. Pomeau, \textit{Soft Matter}, 2011, \textbf{7}, 10612-10619.
\bibitem{Chakrabarti13} A. Chakrabarti and M. K. Chaudhury, \textit{Langmuir}, 2013, \textbf{29},
6926-6935.
\bibitem{Henann14} D. L. Henann and K. Bertoldi, \textit{Soft Matter}, 2014, \textbf{10}, 709-717.
\bibitem{Style12}  R. W. Style and E. R. Dufresne, \textit{Soft Matter}, 2012, \textbf{8}, 7177-7184.
\bibitem{StylePRL13} R. W. Style, Y. Che, J. S. Wettlaufer, L. A. Wilen and
E. R. Dufresne, \textit{Phys. Rev. Lett.}, 2013, \textbf{110}, 066103.
\bibitem{StylePNAS13} R. W. Style, Y. Che, S. J. Park, B. M. Weon, J. H. Je,
C. Hyland, G. K. German, M. P. Power, L. A. Wilen, J. S. Wettlaufer and E. R. Dufresne, \textit{Proc. Natl. Acad. Sci. U. S. A.}, 2013, \textbf{110}, 12541-12544.
\bibitem{Nadermann13}  N. Nadermann, C.-Y. Hui and A. Jagota, \textit{Proc. Natl. Acad. Sci.
U. S. A.}, 2013, \textbf{110}, 10541-10545.
\bibitem{Bostwick14}  J. B. Bostwick, M. Shearer and K. E. Daniels, \textit{Soft Matter}, 2014, \textbf{10}, 7361-7369.
\bibitem{Karpitschka14} S. Karpitschka,	S. Das,	M. van Gorcum,	H. Perrin,	B. Andreotti and J. H. Snoeijer,
\textit{Nat. Commun.}, 2015, \textbf{6}:7891, DOI: 10.1038/ncomms8891.
\bibitem{StyleNatComm13}  R. W. Style, C. Hyland, R. Boltyanskiy, J. S. Wettlaufer and
E. R. Dufresne, \textit{Nat. Commun.}, 2013, \textbf{4}:2728, DOI: 10.1038/ncomms3728.
\bibitem{Salez13} T. Salez, M. Benzaquen and E. Raphael, \textit{Soft Matter}, 2013, \textbf{9}, 10699-10704.
\bibitem{Xu14} X. Xu, A. Jagota and C.-Y. Hui, \textit{Soft Matter}, 2014, \textbf{10}, 4625-4632.
\bibitem{Cao14}  Z. Cao, M. J. Stevens and A. V. Dobrynin, \textit{Macromolecules}, 2014, \textbf{47}, 3203-3209.
\bibitem{Mora13} S. Mora, C. Maurini, T. Phou, J. -M. Fromental, B. Audoly and Y. Pomeau, 
\textit{Phys. Rev. Lett.}, 2013, \textbf{111}, 114301.
\bibitem{Duan07}  H. L. Duan, X. Yi, Z. P. Huang and J. Wang,
\textit{Mech. Mater.}, 2007, \textbf{39}, 81-93.
\bibitem{Brisardbulk10} S. Brisard, L. Dormieux and D. Kondo, 
\textit{Comput. Mater. Sci.}, 2010, \textbf{48}, 589-€"596.
\bibitem{Brisard10} S. Brisard, L. Dormieux, D. Kondo, 
\textit{Comput. Mater. Sci.}, 2010, \textbf{50}, 403-410.
\bibitem{Duan05} H. L. Duan, J. Wang, Z. P. Huang and B. L. Karihaloo, 
\textit{Proc. R.
Soc. A}, 2005, \textbf{461}, 3335-3353.
\bibitem{Duan07secondo} H. L. Duan, X. Yi, Z. P. Huang and J. Wang, 
\textit{Mech. Mater.}, 2007, \textbf{39}, 94-103.
\bibitem{Stylesoft15} R. W. Style, J. S. Wettlaufer and E. R. Dufresne, 
\textit{Soft Matter}, 2015, \textbf{11}, 672-679.
\bibitem{Style15} R. W. Style,	R. Boltyanskiy,	B. Allen, K. E. Jensen,	H. P. Foote,	J. S. Wettlaufer and E. R. Dufresne, 
\textit{Nature Physics}, 2015, \textbf{11}, 82-87.
\bibitem{MSWa} F. Mancarella, R. W. Style, J. S. Wettlaufer, \textit{Interfacial tension and the Mori-Tanaka theory of non-dilute soft composite solids}, subjudice (2015). 
\bibitem{Duan05JMPS} H. L. Duan, J. Wang, Z. P. Huang and B. L. Karihaloo, 
\textit{J. Mech. Phys. Solids}, 2005, \textbf{53}, 1574-1596.
\bibitem{Lur'e64} A. I. Lur'e, \textit{Three-dimensional problems of the theory of elasticity} (translated from the Russian), Interscience, New York, 1964, ch.6, pp. 325-379.
\bibitem{Eshelby56} J. D. Eshelby, \textit{Solid State Physics}, 1956, \textbf{3}, 79-144.
\bibitem{Eshelby57}  J. D. Eshelby, 
\textit{Proc. R. Soc. Lond. A}, 1957, \textbf{241}, 376-396.
\bibitem{Forterre} J. Dumais and Y. Forterre, 
\textit{Annual Review of Fluid Mechanics}, 2012, \textbf{44}, 453-478.
\bibitem{Moeendarbary} E. Moeendarbary, L. Valon, M. Fritzsche, A. R. Harris, D. A.
Moulding, A. J. Thrasher, E. Stride, L. Mahadevan and G. T. Charras, 
\textit{Nature Materials}, 2013, \textbf{12}, 253-261.
\bibitem{Hesse} M. A. Hesse, A. R. Schiemenz, Y. Liang and E. M. Parmentier, 
\textit{Geophysical Journal International}, 2011, \textbf{187}, 1057-1075.
\bibitem{MacMinn:2012} M. L. Szulczewski, C. W. MacMinn, H. J. Herzog and R. Juanes, 
 \textit{Proc. Natl. Acad. Sci. U. S. A.}, 2012, \textbf{109}, 5185-5189.
\bibitem{Wang} H. F. Wang, \textit{Theory of Linear Poroelasticity}, Princeton University
Press, Princeton NJ, 2000.
\bibitem{MacMinn:2015} C. W. MacMinn, E. R. Dufresne and J. S. Wettlaufer, 
\textit{Phys. Rev. X}, 2015, \textbf{5}, 011020.
\bibitem{Smith74} J. C. Smith, 
\textit{J. Res. Nat. Bur. Stand. A}, 1974, \textbf{78} , 355-361.
\bibitem{Smith75} J. C. Smith, 
\textit{J. Res. Nat. Bur. Stand. A}, 1975, \textbf{79}, 419-423.
\end{thebibliography}


\end{document}